\documentclass[preprint,aps,pre,amsmath,amssymb,amsfonts]{revtex4}
\usepackage{mathrsfs}
\usepackage{amsfonts}
\usepackage{amstext}
\usepackage{amsmath}
\usepackage{amssymb}
\newtheorem{theorem}{Theorem}

\newtheorem{cor}[theorem]{Corollary}

\def\be{\begin{equation}}
\def\ee{\end{equation}}
\def\ba{\begin{array}}
\def\ea{\end{array}}

\begin{document}

\title{A Note on Fully Entangled Fraction}
\author{Ming-Jing Zhao$^{1}$}
\author{Zong-Guo Li$^{2}$}
\author{Shao-Ming Fei$^{1}$}
\author{Zhi-Xi Wang$^{1}$}

\affiliation{$^1$School of Mathematical Sciences, Capital Normal
University, Beijing 100048,
China\\
$^2$Beijing National Laboratory for Condensed Matter Physics,
Institute of Physics, Chinese Academy of Sciences, Beijing 100190,
China}

\begin{abstract}
{We investigate the general characters of fully entangled fraction
for quantum states. The fully entangled fraction of Isotropic states
and Werner states are analytically computed.}
\end{abstract}

\pacs{03.67.Mn, 03.65.Ud, 03.65.Yz}

\maketitle
\section{Introduction}

Entanglement is a vital resource for some practical applications in
quantum information processing such as quantum cryptography, quantum
teleportation and quantum computation \cite{bennett2000,nielsen}.
One way to characterize the nonclassical property of quantum
entanglement is to quantify the entanglement in terms of some
measures, for example, entanglement of formation \cite{be4},
concurrence \cite{con}, negativity \cite{G.Vidal} and geometric
measure \cite{Brody, t-c-wei}. However, in fact it is the fully
entangled fraction (FEF) that is tightly related to many quantum
information processing such as dense coding \cite{dc}, teleportation
\cite{tel}, entanglement swapping \cite{es}, and quantum
cryptography (Bell inequalities) \cite{crypto}. For instance the
fidelity of optimal teleportation is given by FEF \cite{bennett,
fefandtel,alb2002}. Additionally, the FEF in two-qubit system acts
as an index to characterize the nonlocal correlation \cite{zwzhou}
and one can never determine whether a state is entangled or not
through the D\"{u}r-Cirac method \cite{Y.Ota}, which is a simple and
effective method for examining multiqubit entanglement, if the FEF
is less than or equal to $\frac{1}{2}$. FEF also plays a significant
role in deriving two bounds on the damping rates of the dissipative
channel \cite{SKO}. Since FEF has a clear experimental meaning, an
analytic formula for FEF is of great importance. In
\cite{grondalski} an elegant formula for FEF in two-qubit system is
derived analytically by using the method of Lagrange multiplier. For
high dimensional quantum states the analytical computation of FEF
remains formidable and less results have been known. In
\cite{upperbound} the upper bound of FEF has been estimated.

In this paper, we first present some properties of FEF and its
relations with negativity, concurrence and geometric measure. Then
we analytically solve the FEF for some classes of quantum states
such as Isotropic states and Werner states.

\section{Properties of FEF}

The FEF of a density matrix $\rho$ in $d\otimes d$ Hilbert space is
defined by \cite{fefandtel,alb2002}
\begin{eqnarray}\label{1}
\mathcal{F}(\rho)&=&\max_U\langle \psi^+|U^{\dagger} \otimes I \rho
U \otimes I |\psi^+\rangle,
\end{eqnarray}
where $U$ (resp. $I$) is a unitary (resp. identity) matrix,
$|\psi^+\rangle = \frac{1}{\sqrt{d}} \sum_{k=1}^d |kk\rangle$ is the
maximally entangled pure state.

Any $d\otimes d$ pure state $|\psi\rangle =\sum_{i,j=1}^d a_{ij} |ij
\rangle$ can be written in the standard Schmidt form,
\begin{eqnarray}\label{ddpure}
|\psi\rangle=\sum_i \lambda_i|ii\rangle,
\end{eqnarray}
where the Schmidt coefficients $\lambda_i$, $i=1,\cdots,d$, satisfy
$0\leq \lambda_d \leq \cdots \leq \lambda_2 \leq \lambda_1 \leq1$
and $\sum_i \lambda_i^2=1$. The FEF of $|\psi\rangle$ has been given in \cite{fp},
\begin{eqnarray}\label{ddpuref}
\mathcal{F}(|\psi\rangle)=\frac{1}{d}(\sum_i \lambda_i)^2.
\end{eqnarray}
From Eq. (\ref{ddpuref}) it can be seen that $|\psi\rangle$ is
separable if and only if $\mathcal{F}(|\psi\rangle)=\frac{1}{d}$.

For pure states the FEF has direct relations with some entanglement
measures. For instance, due to $\|(|\psi\rangle \langle
\psi|)^{T_1}\|=(\sum_i \lambda_i)^2$, the negativity \cite{G.Vidal},
$\mathcal{N}(\rho)= \frac {\| \rho^{T_1}\|-1}{2}$ can be expressed
as
$\mathcal{N}(|\psi\rangle)=\frac{d\mathcal{F}(|\psi\rangle)-1}{2}$,
where $T_1$ stands for partial transposition with respect to the
first space. The geometric measure \cite{t-c-wei} is defined by
$\mathcal{E}(|\psi\rangle)=1- \Lambda_{\max}^2(| \psi \rangle)$,
where $\Lambda_{\max}^2(| \psi \rangle) =\sup_{|\phi\rangle \in S}
|\langle \psi| \phi \rangle|^2$ and $S$ denotes the set of product
states. For pure state $|\psi\rangle$ in Eq. (\ref{ddpure}), we have
$\Lambda_{\max}^2(| \psi \rangle)=\lambda_1^2$ and
$\mathcal{E}(|\psi\rangle)=1- \lambda_1^2$. From Eq.
(\ref{ddpuref}), we can get the relation between FEF and geometric
measure: $d \Lambda_{\max}^2\geq \mathcal{F}$ and $\mathcal{F}\leq d
(1-\mathcal{E})$.

For $d \otimes d$ mixed states $\rho=\sum_i p_i |\psi_i\rangle
\langle \psi_i|$, $\mathcal{F}(\rho)$ has no general analytical
formula. It can be shown that \be\label{t5} \mathcal{F}(\rho)\leq
\sum_i p_i\mathcal{F}(|\psi_i\rangle),
\ee
since
$$
\begin{array}{rcl}
\mathcal{F}(\rho)
\leq \displaystyle\sum_i p_i \max_{U_i} \langle \psi^+|U_i^\dagger
\otimes I |\psi_i\rangle \langle \psi_i| U_i\otimes I |\psi^+\rangle
= \sum_i p_i \mathcal{F}(|\psi_i\rangle).
\end{array}
$$
From Eq. (\ref{t5}) and the main result in \cite{kc2005}, we can
obtain a relation between FEF and concurrence for mixed states,
$\mathcal{C}(\rho)\geq \max
\{\sqrt{\frac{2}{d(d-1)}}(d\mathcal{F}(\rho)-1), 0\}$. For two-qubit
states, using the relation between the entanglement of formation and
the concurrence, one gets the relation between the entanglement of
formation and FEF presented in \cite{bennett}.

Most of the entanglement measures for a mixed state $\rho$ are
defined in terms of all possible pure state decompositions of $\rho$
by convex roof, e.g. for concurrence $C$, $C(\rho) \equiv
\min_{\{p_i, |\psi _i \rangle \}} \sum_i p_i C(|\psi _i\rangle)$. A
question one may ask is whether the FEF of a mixed state also has
such property: $\mathcal{F}(\rho) \equiv \min _{\{p_i, |\psi _i
\rangle \}} \sum_i p_i \mathcal{F}(|\psi _i \rangle)$. The answer is
no. As a counter-example one may consider $2\otimes 2$ state $\rho=
\frac{1}{2} (|00\rangle \langle 00| + |11\rangle \langle 11|)$. By
direct calculation one has $\mathcal{F}(\rho)= \frac{1}{2}$. While
for any other decompositions $\{ p_i, |\psi_i \rangle \}$ with
$|\psi_i \rangle = \alpha_i|00 \rangle +\beta_i |11\rangle$, where
$\alpha_i, \beta_i \in \mathbb{C}$ and $|\alpha_i|^2
+|\beta_i|^2=1$, $\sum_i p_i \mathcal{F} (|\psi_i \rangle)
=\frac{1}{2}+ \sum_i p_i |\alpha_i \beta_i| > \mathcal{F}(\rho)$.
Here we give a condition such that the equality holds in Eq.
(\ref{t5}).

\begin{theorem}
For any $d \otimes d$ mixed state $\rho= \sum_{t=1}^n p_t
|\psi_t\rangle \langle \psi_t|$, $n>1$, $\mathcal{F}(\rho) = \sum_t
p_t \mathcal{F}(|\psi _t \rangle)$ if and only if there exist
unitary transformations $U_1^{(t)}$ and $U_2^{(t)}$ such that
$U_1^{(t)} \otimes U_2^{(t)} |\psi_t \rangle = \sum_j a_j^{(t)}
|jj\rangle$ with $a_j^{(t)} \geq 0$ and $U_1^{(s)\dagger} U_2^{(s)*}
= e^{i\theta_{st}} U_1^{(t)\dagger} U_2^{(t)*}$, $1\leq s, t \leq
n$, $0\leq \theta_{st}\leq 2\pi$. For such state, $\mathcal{F}(\rho)
= \frac{1}{d}\sum_t p_t (\sum_j a_j^{(t)})^2$.
\end{theorem}

{\bf Proof.} We only need to prove the case $n=2$. The cases
$n\geq3$ can be similarly proved.

Assume $\rho=p_1 |\psi_1 \rangle \langle \psi_1| +p_2 |\psi_2
\rangle \langle \psi_2| $. By Schmidt decomposition, there exist
unitary matrices $U_1^{(1)}, U_2^{(1)},  U_1^{(2)}, U_2^{(2)}$ such
that $|\tilde \psi_i\rangle= U_1^{(i)} \otimes U_2^{(i)} |\psi_i
\rangle = \sum_j a_j^{(i)} |jj\rangle $ with $a_j^{(i)} \geq 0$,
$i=1, 2$. We have
\begin{eqnarray*}
\mathcal{F}(\rho)
&=& \max_{V} (p_1 \langle \psi^+| V^\dagger U_1^{(1)\dagger} \otimes
U_2^{(1)\dagger} |\tilde \psi_1 \rangle \langle \tilde \psi_1|
U_1^{(1)} V \otimes  U_2^{(1)} |\psi^+\rangle\\\nonumber &&+ p_2
\langle \psi^+| V^\dagger U_1^{(2)\dagger} \otimes  U_2^{(1)\dagger}
|\tilde \psi_2 \rangle \langle \tilde \psi_2|  U_1^{(2)} V \otimes
U_2^{(2)} |\psi^+\rangle ).\nonumber
\end{eqnarray*}
Therefore $\mathcal{F}(\rho) = p_1 \mathcal{F}(|\psi _1 \rangle)+
p_2 \mathcal{F}(|\psi _2 \rangle)$ if and only if there exists
unitary matrix $V$ such that
$$
\begin{array}{rcl}
\mathcal{F}(|\psi_1 \rangle)&=&\mathcal{F}(|\tilde \psi _1 \rangle)
\\
&=& \langle \psi^+| V^\dagger U_1^{(1)\dagger} \otimes
U_2^{(1)\dagger} |\tilde \psi_1 \rangle \langle \tilde \psi_1|
U_1^{(1)} V \otimes  U_2^{(1)}
|\psi^+\rangle\\
&=& tr (V^\dagger U_1^{(1)\dagger} \otimes
 U_2^{(1)\dagger} |\tilde \psi_1 \rangle \langle \tilde
\psi_1| U_1^{(1)} V \otimes  U_2^{(1)} P_+)\\
&=& tr (  U_2^{(1)*}
 V^\dagger  U_1^{(1)\dagger} \otimes I |\tilde \psi_1
\rangle \langle \tilde \psi_1|  U_1^{(1)} V  U_2^{(1)T} \otimes I
P_+)  \\
&=& |\langle \tilde \psi_1|  U_1^{(1)} V U_2^{(1)T} \otimes
I|\psi^+\rangle|^2,
\end{array}
$$
where $P_+ = |\psi^+\rangle \langle \psi^+|$ and $A \otimes I P_+ =
I \otimes A^T P_+$. Furthermore, $\mathcal{F}(|\psi _2 \rangle)=
\mathcal{F}(|\tilde \psi _2\rangle)= |\langle \tilde \psi_2|
U_1^{(2)} V U_2^{(2)T} \otimes I|\psi^+\rangle|^2 $. On the other
hand, $\mathcal{F}(|\tilde \psi_1\rangle)= \frac{1}{d}(\sum_j
a_j^{(1)})^2$ and $\mathcal{F}(|\tilde \psi_2\rangle)=
\frac{1}{d}(\sum_j a_j^{(2)})^2$. $\mathcal{F}(|\psi _1 \rangle)$
reaches maximum when $U_1^{(1)} V  U_2^{(1)T} = e^{i\theta_1} I$,
i.e. $ U_1^{(1)\dagger} U_2^{(1)*} = e^{-i\theta_1} V $. Similarly,
we have $ U_1^{(2)\dagger} U_2^{(2)*} =e^{-i\theta_2} V$ and
$U_1^{(1)\dagger} U_2^{(1)*} = e^{i(\theta_2-\theta_1)}
U_1^{(2)\dagger}  U_2^{(2)*}$. The value of FEF can be obtained from
Eq. (\ref{ddpuref}). \hfill $\Box$

Theorem $1$ gives the condition that FEF fulfills the convex roof
measure. Besides if one interprets FEF of a state $\rho$ as the
distance between $\rho$ and maximally entangled states, then the
larger FEF is, the closer they are. Although there are infinite
maximally entangled states, the one $U\otimes I |\psi^+\rangle$
which reaches the maximum of Eq. (\ref{1}) is the closest maximally
entangled state to $\rho$. The theorem $1$ also tells us when the
closest maximally entangled state to two different pure states are
the same. As an example, we consider mixed state $\rho=\sum_{i=1}^d
p_i |i, \sigma(i)\rangle \langle i, \sigma(i)|$, where $\sigma$
denotes the permutation of $(1,2,\cdots,d)$. For this state, theorem
applies and we have $\mathcal{F}(\rho)= \sum_{i=1}^d p_i
\mathcal{F}(|\psi_i\rangle)=\frac{1}{d}$ with $|\psi_i\rangle = |i,
\sigma(i)\rangle$. The distance between $|\psi_i\rangle$ and
maximally entangled states is $\frac{1}{d}$. The closest maximally
entangled state to $|\psi_i\rangle$ is
$|\psi_0\rangle=\frac{1}{\sqrt{d}}\sum_{i=1}^d |i,
\sigma(i)\rangle$: $|\langle \psi_i |\psi_0 \rangle|^2
=\frac{1}{d}$, $i=1, \cdots, d$.

From Eq. (\ref{ddpuref}) and Eq. (\ref{t5}) one can obtain that for any
$d \otimes d$ mixed state $\rho$, if $\rho$ is separable,
then $\mathcal{F}(\rho) \leq \frac{1}{d}$. Moreover

\begin{theorem}
For any $d \otimes d$ mixed state $\rho$, $\frac{1}{d^2} \leq
\mathcal{F}(\rho) \leq 1$. $\mathcal{F}(\rho)=1$ if and only if
$\rho$ is a maximally entangled pure state.
$\mathcal{F}(\rho)=\frac{1}{d^2}$ if and only if $\rho$ is the
maximally mixed state, i.e. $\rho=\frac{1}{d^2}I$.
\end{theorem}

{\bf Proof.}  For any $d \otimes d$ mixed state $\rho$, we assume
$\rho=\sum_{i=1}^{d^2} \lambda_i |\phi_i\rangle \langle \phi_i|$ is
the spectrum decomposition such that $\sum_{i=1}^{d^2}\lambda_i=1$,
$0 \leq \lambda_i \leq 1$ and $\{ |\phi_i\rangle \}_{i=1}^{d^2}$ are
normalized orthogonal eigenvectors in $d \otimes d$ Hilbert space.
Then $\mathcal{F}(\rho)={\displaystyle\max_U}\langle
\psi^+|U^\dagger \otimes I \rho U\otimes I |\psi^+\rangle =
{\displaystyle\max_U} \sum_i \lambda_i \langle \psi^+|U^\dagger
\otimes I |\phi_i\rangle \langle \phi_i| U\otimes I |\psi^+\rangle$.
Set $a_i = \langle \psi^+|U^\dagger \otimes I |\phi_i\rangle \langle
\phi_i| U\otimes I |\psi^+\rangle$, which satisfies $0 \leq a_i \leq
1$ and $\sum_{i=1}^{d^2} a_i =1$ due to the completeness of the
eigenvectors $\{ |\phi_i\rangle \}$. $\sum_{i=1}^{d^2} \lambda_i a_i
\leq \sum_{i=1}^{d^2} \lambda_i =1$ becomes an equality if and only
if there are only one nonzero coefficient, say, $a_i=1$ and one
nonzero coefficient $\lambda_i=1$. Therefore $\mathcal{F}(\rho)=1$
if and only if $\rho$ is maximally entangled pure state.

On the other hand, the minimum of the function $g(\lambda_i,
a_i)=\sum_{i=1}^{d^2} \lambda_i a_i$ is $\frac{1}{d^2}$ by Lagrange
multiplier. It reaches its minimum if and only if $\lambda_i=
a_i=\frac{1}{d^2}$ for $i=1, \cdots, d^2$. This gives rise to
$\rho=\frac{1}{d^2}I$. \hfill $\Box$

Similar to the proof above, here one can also obtain the range of
geometric measure for mixed states.

\begin{cor}
For any $d \otimes d$ mixed state $\rho$, it satisfies $0 \leq
\mathcal{E}(\rho) \leq \frac{d-1}{d}$. $\mathcal{E}(\rho) =0$ if and
only if $\rho$ is a separable state. $\mathcal{E}(\rho)
=\frac{d-1}{d}$ if and only if $\rho$ is a maximally entangled pure state.
\end{cor}

We have studied some properties related to the FEF. Before we
compute analytically the FEF for Isotropic states and Werner states,
we investigate another property that similarly studied for
entanglement of formation, negativity, concurrence, geometric
measure and q-squashed entanglement \cite{superposition}.

\begin{theorem}
For two given pure states $|\phi_1\rangle$ and $|\phi_2\rangle$, the
FEF of their superposition $|\psi\rangle = \frac{1}{\gamma} (\alpha
|\phi_1\rangle + \beta |\phi_2\rangle )$ satisfies:
\begin{eqnarray}\label{superposition}
\max\{ ||\alpha| \mathcal{F}^{\frac{1}{2}}(|\phi_1\rangle)-|\beta|
\mathcal{F}^{\frac{1}{2}}(|\phi_2\rangle)|, \frac{1}{d^2} \} \leq
|\gamma| \mathcal{F}^{\frac{1}{2}}(|\psi\rangle) \leq \min\{|\alpha|
\mathcal{F}^{\frac{1}{2}}(|\phi_1\rangle)+|\beta|
\mathcal{F}^{\frac{1}{2}}(|\phi_2\rangle), 1\}.
\end{eqnarray}
\end{theorem}

{\bf Proof.} By the definition of FEF we have
\begin{eqnarray*}
\mathcal{F}(|\psi\rangle)&=&\frac{1}{\gamma^2} \max_U \langle
\psi^+| U^\dagger \otimes I (\alpha |\phi_1\rangle + \beta
|\phi_2\rangle) (\alpha^* \langle \phi_1| +\beta^* \langle \phi_2|)
U \otimes I
|\psi^+\rangle\\
&\leq& \frac{1}{\gamma^2} (|\alpha|^2 \mathcal{F}(|\phi_1\rangle) +
|\beta|^2 \mathcal{F}(|\phi_2\rangle) + 2|\alpha
\beta|\sqrt{\mathcal{F}(|\phi_1\rangle)\mathcal{F}(|\phi_2\rangle)})\\
&=& \frac{1}{\gamma^2} (|\alpha|
\mathcal{F}^{\frac{1}{2}}(|\phi_1\rangle)+|\beta|
\mathcal{F}^{\frac{1}{2}}(|\phi_2\rangle))^2,
\end{eqnarray*}
which gives the right hand side of Eq. (\ref{superposition}).

Similarly, taking into account of $|\phi_1\rangle =
\frac{\gamma}{\alpha} |\psi\rangle-
\frac{\beta}{\alpha}|\phi_2\rangle$ and $|\phi_2\rangle =
\frac{\gamma}{\beta} |\psi\rangle-
\frac{\alpha}{\beta}|\phi_1\rangle$, one gets the left hand side of Eq.
(\ref{superposition}).  \hfill $\Box$

For example, let $|\phi_1\rangle = |00\rangle$, $|\phi_2\rangle =
|11\rangle$ and $|\psi\rangle = \frac{1}{\sqrt{2}} (|00\rangle
+|11\rangle)$, then FEF of $|\psi\rangle$ reaches the upper bound of
Eq. (\ref{superposition}). If we take $|\phi_1\rangle =
\frac{1}{\sqrt{2}} (|00\rangle -|11\rangle)$, $|\phi_2\rangle =
|11\rangle$ and $|\psi\rangle = |00\rangle$, then FEF of
$|\psi\rangle$ reaches the lower bound of Eq. (\ref{superposition}).
Eq. (\ref{superposition}) can also be generalized to the case of
superposition with more than two components: for $|\psi\rangle=
\frac{1}{\gamma}(\alpha_1 |\phi_1\rangle +\cdots + \alpha_m |\phi_m
\rangle)$, we have $\max_i\{ |\alpha_i|
\mathcal{F}^{\frac{1}{2}}(|\phi_i\rangle)- \sum_{j\neq i} |\alpha_j|
\mathcal{F}^{\frac{1}{2}}(|\phi_j\rangle), \frac{1}{d^2}\} \leq
|\gamma| \mathcal{F}^{\frac{1}{2}}(|\psi\rangle) \leq \min\{\sum_{i}
|\alpha_i| \mathcal{F}^{\frac{1}{2}}(|\phi_i\rangle),1\}$.

\section{FEF for some classes of mixed states}

Generally for mixed states it is rather difficult to get analytical
formulae for entanglement measures and FEF. Nevertheless for some
special quantum states, elegant results have been derived. For
instance, for the Isotropic state, entanglement of formation
\cite{Barbara}, concurrence \cite{Rungta} and geometric measure
\cite{t-c-wei} have been calculated explicitly. For the Werner
state, concurrence \cite{kc2006} and geometric measure
\cite{t-c-wei} have been investigated also. Now we calculate
analytically FEF for such well-known mixed states.

\emph{Isotropic state}
~Isotropic states \cite{fp} are a class of $U \otimes
U^*$ invariant mixed states in $d \otimes d$ Hilbert space:
\begin{eqnarray}
\rho_{iso}(f)
&=&\frac{1-f}{d^2-1} I +  \frac{d^2f-1}{d^2-1} |\psi^+ \rangle
\langle \psi^+|,
\end{eqnarray}
with $f=\langle \psi^+| \rho_{iso}(f) |\psi^+ \rangle$ satisfying
$0\leq f \leq 1$. These states are shown to be separable if and only
if they are PPT, i.e. $f \leq \frac{1}{d}$. They can be distilled if
they are entangled, which means $f> \frac{1}{d}$ \cite{fp}.

By definition, the FEF is given by
$$
\mathcal{F}(\rho_{iso}(f))
=\frac{1-f}{d^2-1} + \max_U \frac{d^2f-1}{d^2-1} |\langle \psi^+|
U\otimes I |\psi^+ \rangle |^2 = \frac{1-f}{d^2-1} + \max_U
\frac{d^2f-1}{d^2-1} |\frac{1}{d} tr U|^2.
$$
If $\frac{d^2f-1}{d^2-1}
>0$, i.e. $f> \frac{1}{d^2}$, we have $\mathcal{F}(\rho_{iso}(f))=\frac{1-f}{d^2-1} +
\frac{d^2f-1}{d^2-1} = f$. The maximum is attained by choosing
$U=I$. If $\frac{d^2f-1}{d^2-1} <0$, i.e. $f< \frac{1}{d^2}$, we get
$\mathcal{F}(\rho_{iso}(f))=\frac{1-f}{d^2-1} + \frac{d^2f-1}{d^2-1}
{\displaystyle\min_U |\frac{1}{d} tr U|^2}\leq \frac{1-f}{d^2-1}$.
In fact, if we choose $U= \sum_{i \neq j} |i\rangle \langle j|$,
then the inequality becomes an equality. If $\frac{d^2f-1}{d^2-1}
=0$, i.e. $f= \frac{1}{d^2}$, we have $\mathcal{F}(\rho_{iso}(f))=
\frac{1}{d^2}$. Therefore we get the FEF for Isotropic states:
$$
\mathcal{F}(\rho_{iso}(f))=\left\{
   \begin{array}{cc}\displaystyle
   f, & \frac{1}{d^2} \leq f \leq 1;\\
\displaystyle  \frac{1-f}{d^2-1}, & 0 \leq f<\frac{1}{d^2}.
   \end{array}\right.
$$

According to \cite{fefandtel}, the fidelity $\bf{f}_{\max}$ of
optimal teleportation via state $\rho$ attainable by means of
trace-preserving local quantum operations and classical communication (LOCC) is equal to ${\bf
f}_{\max}(\rho)=\frac{\mathcal{F}(\rho) d +1}{d+1}$. If
$\mathcal{F}(\rho)> \frac{1}{d}$, then state $\rho$ is said to be useful for
teleportation. Hence all entangled Isotropic states are useful in quantum teleportation.

\emph{Werner state}
~Werner states \cite{werner} are a class of $U \otimes U$ invariant
mixed states in $d \otimes d$ Hilbert space:
\begin{eqnarray}
\rho_{wer}(f) = \frac{d-f}{d^3-d} I + \frac{df-1}{d^3-d}V,
\end{eqnarray}
where $V=\sum_{i,j=1}^d |ij\rangle \langle ji|$ and $f=\langle
\psi^+| \rho_{wer}(f) |\psi^+ \rangle$, $-1\leq f \leq 1$. These
states are shown to be separable if and only if they are PPT ($f
\geq 0$).

The FEF of Werner state is given by
$$\displaystyle
\ba{rcl}
\mathcal{F}(\rho_{wer}(f))
&=&\displaystyle\frac{d-f}{d^3-d} + \max_U \frac{df-1}{d^3-d} |\langle \psi^+|
U^\dagger \otimes I V U \otimes I |\psi^+ \rangle |\\[3mm]\displaystyle
&=&\displaystyle\frac{d-f}{d^3-d} + \max_U \frac{df-1}{d^4-d^2}
\sum_{kl} \langle k|U^\dagger|
l\rangle \langle k|U| l\rangle\\[3mm]\displaystyle
&=&\displaystyle\frac{d-f}{d^3-d} + \max_U \frac{df-1}{d^4-d^2}
tr(UU^*). \ea
$$

i) If $df-1 >0$, since $UU^*$ is unitary,
$$\mathcal{F}(\rho_{wer}(f))= \frac{d-f}{d^3-d} +
\frac{df-1}{d^3-d} = \frac{f+1}{d(d+1)},$$
which corresponds to the case $U=I$.

ii) If $df-1 <0$ and $d$ is even, we get
$$\mathcal{F}(\rho_{wer}(f))=\frac{d-f}{d^3-d} -
\frac{df-1}{d^3-d} = \frac{1-f}{d(d-1)},$$
which can be attained by choosing $U=
A_{2\times 2} \otimes I_{\frac{d}{2} \times \frac{d}{2}}$ with
$A_{2\times 2}=\left(
\begin{array}{cc}
0 & 1\\
-1 & 0 \end{array} \right)$.

iii) For the case of $df-1< 0$ and $d$ is odd, one has
$$\mathcal{F}(\rho_{wer}(f))=\frac{d-f}{d^3-d} + \frac{df-1}{d^3-d} \times
\frac{-d+2}{d} = \frac{d^2-d^2f+df+d-2}{d^2(d^2-1)}.$$

vi) If $df-1 =0$, i.e. $f= \frac{1}{d}$, $\mathcal{F}(\rho_{wer}(f))= \frac{1}{d^2}$.

Therefore we get the FEF for Werner states:
$$ \mathcal{F}(\rho_{wer}(f))=\left\{
   \begin{array}{cc}\displaystyle
   \frac{f+1}{d(d+1)}, & \frac{1}{d} \leq f \leq 1;\\\displaystyle
 \frac{1-f}{d(d-1)}, & -1 \leq f<\frac{1}{d}.
   \end{array}\right.
   $$
if d is even; and
$$ \mathcal{F}(\rho_{wer}(f))=\left\{
   \begin{array}{cc}\displaystyle
   \frac{f+1}{d(d+1)}, & \frac{1}{d} \leq f \leq 1;\\\displaystyle
 \frac{d^2-d^2f+df+d-2}{d^2(d^2-1)}, & -1 \leq f<\frac{1}{d}.
   \end{array}\right.$$
if d is odd. Hence this formula tells us there exist entangled
Werner states which are not useful for teleportation.

\section{Conclusions}
We have explored some characters of FEF and analytically computed
the FEF of several well-known classes of quantum mixed states. These
results complement previous ones in this subject and may give rise
to new application to the quantum information processing.

\bigskip
\noindent{\bf Acknowledgments}\, This work is supported by the NSF
of China (Grant Nos. 10875081,10871227), the NSF of Beijing (Grant
No. 1092008), KZ200810028013 and PHR201007107.

\end{document}